\documentstyle[preprint,aps]{revtex}
\begin{document}
\preprint{MSU-HEP/50327}
\preprint{hep-ph/9503453}   
\title{Optimized Top Quark Analysis with the Decision Tree}
\author{P. Agrawal, D. Bowser-Chao, and J. Pumplin\cite{byline}
}
\address{
Department of Physics \& Astronomy \\
Michigan State University \\
East Lansing, MI
}
\date{\today}
\maketitle
\begin{abstract}

We present an optimized and physically motivated method for
separating top quark signal events from background events at
the Tevatron.  For the top quark signal $t\bar t \to e/\mu + 4$
jets, we show how to reject all but $25\%$ of the background in
a data sample while retaining $80\%$ of the signal, without
introducing bias into the subsequent mass measurement.  The
technique used is the Binary Decision Tree. Combining this
highly efficient procedure for signal identification with a
novel algorithm for top quark reconstruction, we propose a
powerful new way to measure the top quark mass.

\end{abstract}
\narrowtext

The CDF and D\O\  collaborations recently announced the much-awaited
discovery of the top quark \cite{cdf,d0}.  Both collaborations will
next endeavor to study its production and decay properties further,
and to improve the measurement of its mass.  An important aspect of
the analysis is the need to reject a good fraction of the numerous
background events, while keeping most of the signal.

In this Letter, we employ an artificial-intelligence algorithm,
the Binary Decision Tree \cite{brent}, to discover optimized and
physically motivated cuts that discriminate signal from background
with an efficiency well beyond what is possible using conventional
methods \cite{bop}. By exploiting differences between the signal
and background
without relying on explicit reconstruction of the top quark signal,
these cuts moreover introduce no bias into measurement of the mass.
After presenting the optimized cuts, we propose a new top quark mass
reconstruction algorithm in which a peak in a selected 3-jet mass
distribution reveals $t \to jjj$ and provides a direct measurement of
$m_t$ along with a model-independent
measurement of the background.  With the anticipated integrated
luminosity of the current experimental run at the Tevatron, there
will be enough events not only to see the mass peak clearly, but also
to observe the subsequent hadronic decay $W \to jj$, furnishing a new,
direct
calibration of hadronic calorimetry and the jet-finding algorithm.

In the Standard Model, the top quark decays electroweakly via
$t\to W^+ b$.  The $W$ boson in turn decays hadronically to two jets
($W^+\to jj$) approximately 2/3 of the time, and  semi-leptonically
($W^+\to e^+\nu_e,\mu^+\nu_\mu,\tau^+\nu_\tau$) in the remaining 1/3.
At the Tevatron, top quarks are mainly produced in pairs
$p\bar p\to t\bar t + X$.
Due to severe QCD backgrounds, reliable detection of a top quark
pair requires at least one of the two resulting $W$ bosons to decay
semi-leptonically into $e$ or $\mu$.  We will focus on the
``single leptonic'' signature $\ell + 4$ jets where $\ell=e$ or
$\mu$.  These events occur with six times the rate for double-lepton
events, and have the added virtue of containing only one neutrino,
which facilitates the mass measurement.

The main background to this mode is from the direct production of
$p\bar p\to W + 4$ jets, occurring at about two times the signal rate
in the Standard Model. To suppress this background,
one can exploit the fact that two of the 4 jets in the signal are due
to $b$-quarks which can be tagged with some probability, while
$b$-jets are rare in the background.  Because we seek high signal
acceptance, we will eschew a $b$-tagging requirement, but point out
below how it can be used, when available, to complement our analysis.

In the absence of $b$-tagging, the weapon of choice for reducing the
background is to impose cuts in appropriate observables.   Consider
for example $m_{jj}^6$, whose distribution is shown in Fig.~1.
($m_{jj}^6$ is the lowest of the 6 invariant masses
formed from pairs of the 4 jets.)
The signal peaks near $75$ GeV, while the background (dotted curve)
is concentrated at low $m_{jj}^6$.
Requiring each event to have a minimum observed
$m_{jj}^6$ can thus increase the signal/background ratio $S/B$,
without appreciable loss of signal.

Our first improvement over
previous analyses comes from introducing new variables, including
$m_{jj}^6$, and showing how the physics of the background and signal
makes these variables powerful tools for signal enhancement.  The
major thrust of our work, however, is toward
obtaining cuts in {\it a set of observables} simultaneously.  Before
describing how the Binary Decision Tree determines these highly
efficient cuts, we review the conventional route to
signal {\it vs.} background discrimination.

Based on comparisons of signal and background distributions
like Fig.~1, a list of candidate observables is selected.  A simple
cut specified by  $x_i > x_{i,\min}$
and/or $x_i < x_{i,\max}$ in each variable $x_i$ is arrived at
by trial-and-error adjustment, compromising between background
rejection and signal acceptance.  Each cut is relaxed or tightened
in turn to roughly optimize $S/B$ at the desired level of signal
acceptance. The virtue of this procedure is that the physical nature
of each
cut is understandable.  For example, the simple cut $m_{jj}^6 > 50$ GeV
enhances signal because the background's jets tend to arise from
bremsstrahlung, where the collinear and soft singularities of QCD give
rise to low pair masses.  If there are two or more variables, however,
simple cuts are usually far from optimal. Consider the case of just
two observables. One could examine the two-dimensional
scatter-plot of the signal and background to select an
$S/B$-enhancing cut.  Simple cuts would partition the
scatter-plot along lines running parallel to the coordinate
axes, with events in one or three of the resulting quadrants
to be accepted and all
others rejected. Let us further assume the signal and background
distributions are Gaussian. In this case, an optimal cut is generally
along an ellipse or hyperbola which is the contour of constant $S/B$,
and it cannot be written as one or even several simple cuts.
Even in the special case
where the optimal cut lies along a straight line (which happens
in the Gaussian case when the signal and
background are identical except for their centroids), that line is
generally not
a simple cut, because it need not be parallel to a coordinate axis.
Furthermore, as Fig.~1 shows, the variables used here are obviously
not Gaussian\cite{hmatrix}, so the form of the optimal cut
is not apparent. It is unlikely, however, that the optimal cut is
close to any set of simple cuts. Thus, finding the proper cuts by
hand is difficult for two variables, and seemingly impossible for
more than two variables.

The neural network approach \cite{peterson,neuraltop} offers an
alternative for signal/background classification that avoids the
restrictive form of simple cuts.  It has the unfortunate
drawback, however, of yielding a ``black-box'' solution whose cuts
are not easy to interpret in physical terms.  In addition, the
``training'' of the network to arrive at the final cuts can make
heavy demands on computer time. Some other algorithms that have been
considered, including $H$-matrix and Probability Density
Estimation\cite{hmatrix}, also efficiently separate signal from
background, but fail to match the transparency of simple cuts.

In this paper we advocate instead the
Binary Decision Tree  \cite{brent}, which, compared to the
conventional method, yields much higher signal efficiency. The
decision tree has been shown to perform at the same level as the
neural network in an earlier simple study of the top quark signal
\cite{dzialo}, but with the crucial distinction that it yields
explicit physically interpretable cuts and makes more modest demands
on computer horsepower. The basic  decision tree was described in
Refs.~\cite{brent,dzialo}.  We outline  the algorithm in the form
implemented in the program {\sc hastac} \cite{hastac}, which has been
tailored for use in  high-energy physics signal identification.

Let the set of variables $x = (x_1,\ldots,x_n)$ define the feature
space of events, with each $x_i$ an observable such as $m_{jj}^6$.  A
generalized cut in $(x_1,\ldots,x_n)$ is the requirement that each
event satisfy the inequality $\hat a \cdot (x-x^0) > 0$, where
$\hat a =(\hat a_1,\ldots,\hat a_n)$
is a vector normalized to $\sum_i{\hat a_i^2} = 1$.
The geometrical interpretation of this
expression is clear:  the feature space is cut in two by a
hyperplane passing through the point $x^0$, with the hyperplane
orientation specified by its normal $\hat a$.
Simple single-variable cuts are just hyperplanes
restricted to normals along one coordinate axis of the feature space.
The power of the decision tree derives from its ability to optimally
determine $\hat a$ and $x_0$ for one or more generalized cuts.
In this
paper, we will restrict ourselves to two or three generalized cuts,
which are sufficient to strongly suppress the background.

The optimized hyperplane cuts are found by the decision tree as
follows \cite{brent}.  Approximating the step function as
$\theta(\lambda)\approx
\Theta(\lambda) = (1 + e^{- \lambda / T})^{-1}$, where T is a
relatively small number, the number of signal events $S_{\hbox{A}}$
falling on the ``accepted'' side of the hyperplane is approximated by
\begin{equation}
{S_{\hbox{A}}}(\hat a, x^0) =
\sum_\alpha \Theta(\hat a \cdot(x(\alpha) - x^0))\;,
\end{equation}
with a sum over all signal events $\alpha$.
${B_{\hbox{A}}}(\hat a, x^0)$ is
defined analogously for the background. With ${S_{\hbox{A}}}$ and
${B_{\hbox{A}}}$
thus transformed into
differentiable functions of $\hat a$ and $x^0$, we employ conjugate
gradient optimization \cite{nonlin} to maximize
\begin{equation}
Q_{N}(\hat a, x^0) =
{S_{\hbox{A}}}(\hat a, x^0) /[{B_{\hbox{A}}}(\hat a, x^0)]^N \, .
\end{equation}
The parameter $N$ can be chosen to assign primary importance
to $S/B$ enhancement ($N=1 \Rightarrow Q=S/B$) or to high signal
acceptance ($N \to 0 \Rightarrow Q \to S$).  The value
$N=0.5 \Rightarrow Q=S/\sqrt{B}$ makes the optimized function $Q$
equal to the approximate statistical significance $S/\sigma_B$
of the signal, assuming $S$ and $B$ to be Poisson distributed.
After optimization, each cut is specified by $\{\hat a, x^0\}$, or
more concisely by a form $a \cdot x < c$, where $c$ is a number.
Qualitative interpretation of each cut is through the relative signs
of ${a_i}$, which indicate positive or negative correlation in each
variable with the likelihood of an event being signal.

Next, we describe the physical features of the signal and background
on which our efficient cuts are based.
The primary background to the top quark signature $\ell + 4$
jets is the set of processes leading to direct production of
$W + 4$ jets.  After minimal acceptance cuts given below, about
$40\%$ of the background is due to $q \bar q \to W g g g g $
processes.  The other major sources of
background are $q g \to W g g g q $,
$q q \to W g g q q $ and $q g \to W g q q q$
($q =$ quark or anti-quark), with contributions
ranging from $15 \%$ to $30 \%$.  The background
is thus characterized by processes with multiple gluon jets in the
final states.  The structure of the matrix elements dictates that
much of the cross-section will
lie in regions in phase space close to collinear and/or infrared
divergences.  Near-collinear radiation of jets with respect to the
incoming $p,\bar p$ leads to jets with low transverse momentum $p_T$
(due to quark and gluon bremmstrahlung) and/or high pseudorapidity
$\eta = -\log\tan{\theta/2}$. Collinear and infrared divergences
influence gluon bremmstrahlung and splitting, leading to production
of $q+g$ or $g+g$ with small relative angle and
low dijet mass $m_{jj}$. The trijet masses $m_{jjj}$ similarly
tend to be low.

In strong contrast, the large
mass of the top quark pair implies that it is produced with low
velocity ($50\%$ of the time with $v/c < 0.32\,$) at the Tevatron
energy $\sqrt s = 1.8$ TeV.
( The bulk of top quark pair production is
through $q \bar{q} \to t \bar{t}$, with the next largest contribution
$g g  \to t \bar{t}$ representing only about $10 \%$.)
The velocities of $t$ and $\bar t$ are also small.
The two-body top quark decay $t\to bW^+$ is roughly isotropic in the
top rest frame, giving the $b$ jet a characteristic maximum transverse
momentum scale
$\sim m_t/2$.  In fact, we find, for a top quark of mass $175$
GeV, that the $b$-jet $p_T$ distribution peaks
at $52 \, {\rm GeV}$ with average $71 \, {\rm GeV}$.  The jets from
hadronic $W$ decay share the $W$ momentum, so their average $p_T$ is
somewhat smaller but still peaks at $32 \, {\rm GeV}$ with average
$56 \, {\rm GeV}$.
One expects large trijet masses ($m_{jjj} \sim m_t$), and also large
dijet masses:  $m_{jj} \sim m_W$, or $m_{jj} \sim m_t/\sqrt{3}$ in
view of the kinematic relation
$m_{123}^2 = m_{12}^2 + m_{13}^2 +m_{23}^2$.

Before making a detailed comparison of signal and background, we
list the minimal acceptance cuts we impose to simulate detector
acceptance, and describe our calculation of the signal and background.
The acceptance cuts are
\begin{eqnarray}
 &  &  p_T(j) > 17.0 \, {\rm GeV}, \quad
    p_T(\ell) > 20.0 \, {\rm GeV}, \quad
  \not{\! p_T} > 25.0 \, {\rm GeV}, \quad   \nonumber \\
 &  &  |\eta(j)| < 2.0 , \quad
       |\eta(\ell)| < 2.0,\quad
      R(j,j^\prime) > 0.7, \quad
      R(j, \ell) > 0.4 \, .
\label{eq:accept}
\end{eqnarray}
Here $R(j,j^\prime)=\sqrt{(\Delta \eta)^2+(\Delta\phi)^2}$
where $\Delta\phi$ and $\Delta\eta$ are the differences in
azimuthal angle $\phi$ and pseudorapidity between jets
$j$, $j^\prime$.
To simulate detector resolution, the $\eta$ and $\phi$ of each parton
was smeared from its true value by Gaussian random amounts with
standard deviation $0.05$ in each.
The missing transverse momentum $\not{\! p_T}$, which is taken as a
measurement of $p_T^\nu$, was calculated by smearing each parton
energy by a Gaussian random amount with
$\sigma(E)/E = C/\sqrt{E_T}$ where $C = 0.6$ for jets and $0.15$ for
$\ell$, before calculating the transverse momentum imbalance.  To
simulate the effects of hadronization, we further smeared the
jet energies so that $\sigma(E)/E = 1.0/\sqrt{E_T}$.

We employed helicity amplitude techniques to compute top quark
production, keeping all top quark and $W$ boson decay correlations.
To calculate the background, we used the Monte-Carlo
package {\sc vecbos} \cite{vecbos}.  We used the CTEQ2 set $5$
parton distributions \cite{cteq}, which are leading-order fits and
hence appropriate for our leading-order calculation.  Similarly,
we used a leading-order form for $\alpha_s$, with
$\Lambda_{\rm QCD}$ given by the parton distributions.
Factorization and renormalization scales were chosen as
$\mu_R = \mu_F = m_t$ for the signal and $\mu_R = \mu_F = m_W$
for the background.  The background rate in particular has
theoretical uncertainties, so its direct measurement described below
is most welcome.  We will discuss the specific case of $m_t=175$ GeV
in considerable detail, but also include results for
$m_t=190$ GeV in the table for a
comparison. These results show that the efficacy of both our
cuts and our top mass-reconstruction depend very weakly on the true
value of $m_t$.

Assuming the projected integrated luminosity
$\int {\cal L} dt =100 $pb$^{-1}$ for ``Run I'' at the Tevatron,
we expect a total
of 49 top quark signal events (for $m_t=175$ GeV), and 116 background
events, to pass the minimal acceptance cuts (\ref{eq:accept}).  Thus
we begin with $S/B \sim 0.42$ before our discrimination cuts.

Some variables that we have tried as input to the decision tree
program {\sc hastac} are ordered versions of the observables
discussed above.  The jet transverse momenta are
$p_T^1(j) > \ldots > p_T^4(j)$.  The jet pseudorapidities are
$|\eta^1(j)| > \ldots > |\eta^4(j)|$.  The dijet masses are
$m_{jj}^1 > \ldots > m_{jj}^6$ and the trijet masses are
$m_{jjj}^1 > \ldots > m_{jjj}^4$.

Even before application of the decision tree, several of these
variables point out significant differences between signal
and background.  In the signal, one pair of jets comes from the
decay of a $W$, so the minimum dijet mass $m_{jj}^6$ is less than
$m_W$ except for smearing effects and the $W$ width.
The pair masses otherwise tend to be large, so as shown in Fig.~1,
the signal climbs steadily with $m_{jj}^6$ to a peak near $m_W$,
after which it drops sharply.  In contrast, the
background falls quickly from its largest value at $m_{jj}^6\approx
2 \, p_T^{\min}(j) = 34$ GeV.  A simple cut $m_{jj}^6 > 40$~GeV
passes $(S,B)=(43,50.2)$ events. A tighter cut could even raise
$S/B$ above one:  $m_{jj}^6 > 56$~GeV passes $(S,B)=(30.8, 18.2)$
events. The power of this variable reflects the qualitative
differences between signal and background described above.

Similarly, the distribution in lowest trijet mass $m_{jjj}^4$
for the signal rises to a peak near $m_t$ and then falls sharply
because the signal cannot have a minimum trijet mass above
$m_t$ (modulo jet resolution and width of the top quark).
Meanwhile the background distribution
falls steadily with $m_{jjj}^4$. The simple cut
$m_{jjj}^4 > 120$~GeV would accept $(S,B)=(42.0,44.0)$ events.
However, unlike the $m_{jj}^6$ distribution, a tighter cut would not
yield any further significant enhancement in $S/B$.
It is interesting to note that $m_{jjj}^4$ by itself could serve
as a crude but effective method to directly detect a top
quark mass peak above a smoothly falling background, without
recourse to any fitting procedure or assumptions about the value
of $m_t$.

As expected from the infrared-enhancement in the QCD background, the
minimum jet transverse momentum $p_T^4(j)$ also distinguishes well
between signal and background. The cut $p_T^4(j) > 25$~GeV keeps
$(S,B)=(35.2,36.2)$ events. An extremely tight cut of $p_T^4(j) >
35$~GeV will raise $S/B$ to  more than 2, at the cost of signal
acceptance, with $(S,B)=(17.2,8.2)$ events.  We note in
passing that D\O\  employed a related variable, the scalar sum of the
jet transverse momenta $H_T = \sum_i  |p_T^i(j)|$. A high signal
acceptance cut in $H_T$ is as good as $p_T^4(j)$ --- taking $H_T >
210$ GeV, the
events passing this cut are $(S,B)=(37.4,39.4)$ --- but no amount of
tightening the cut on $H_T$ will obtain $S/B$ significantly over 1.

The three observables just discussed, $m_{jj}^6$, $m_{jjj}^4$, and
$p_T^4(j)$, are the most powerful discriminators we have found, as
judged by their solo performances.  We used them as input to the
{\sc hastac} optimization, in concert with four additional variables
whose individual distributions do not so clearly separate signal
{}from background, but which prove useful in correlation with
the first three.  We should note that several other observables
(\hbox{\it e.g.}{},
the other $m_{jj}^i$, $m_{jjj}^i$, and $p_T^i(j)$) are similarly
helpful, so that our choice of variables was dictated largely by
taste and the ease in interpreting the final cuts.

The first two of these four additional variables are the two largest
jet pseudorapidities $|\eta^4(j)|$ and $|\eta^3(j)|$, which complement
$p_T^4(j)$ in recognizing jet radiation that is collinear with the
incoming beams.  A third  is the $W$ transverse momentum $p_T(W)$,
which
peaks toward $\sim m_t/2$ in the signal and tends to be smaller in
the  background.  The fourth added observable is the maximum trijet
mass $m_{jjj}^1$, which can serve to  close a ``high-mass'' loophole
for the background: though most  dijet masses in the background tend
to be low, some very high  masses can be generated when two gluons,
for example, are radiated  off opposite incoming beams.  Because of
the acceptance cut on $p_T(j)$, there is usually another jet
with sufficient $p_T(j)$ to combine with the pair to produce a large
trijet mass in addition to the large dijet mass.

To follow the conventional route at this point, one would make
cuts in several of these variables simultaneously, and by
trial-and-error adjust the cuts for the best discrimination.
That route would not only be laborious; it would also totally miss any
useful {\it correlations} between the variables, because it permits
only
``rectangular'' cuts.  We therefore presented the variables to {\sc
hastac} for automatic generation of efficient generalized cuts. We
detail our generalized cuts in Table~1. Because two of the cuts
involve variables of different dimension, we scale all
momenta and masses by $m_W$ for convenience. We compare the
background with a signal for $m_t=175$ GeV in the following, but note
that very similar results for $m_t=190$ GeV are indicated in Table~1.

The first generalized
cut (a) drastically shrinks the background by
simultaneously requiring high $p_T^6(j)$, $m_{jjj}^4$, $m_{jj}^6$ and
$p_T(W)$, precisely as anticipated in the discussion above. The
advantage of generalized cuts shows up in the extra $25\%$ decrease of
background relative to cuts in any one variable for the same signal
efficiency. The second cut (b) attempts to close the high-dijet mass
loophole, while at the same time requiring more centrally located jets.
The cuts (a--b) pass $79\%$ of the signal, but only $17\%$ of the
background, giving $S/B$ almost as high as the tight cut in $p_T^4(j)$
described above, but with {\it twice} the signal acceptance. This set
of high-acceptance cuts (a--b) will serve as the starting point for our
reconstruction of the top quark mass. But first we comment on the more
stringent third and fourth cuts.

These two cuts function similarly to cut (b), but have much lower
signal acceptances. They are intended only for the sake of
illustrating how an even higher $S/B$ can be obtained without explicit
top quark reconstruction (though the latter clearly may also be used
to increase $S/B$). Indeed, in the more extreme case (cuts (a--b) and
(d)), the signal/background ratio is almost 4, which is unattainable
through any of the variables taken individually. It is also interesting
to note (bearing in mind that we have not included full hadronization
and detector effects) that the more moderate set of cuts (a--c) is
comparable in both $S/B$ and signal acceptance to that achieved by CDF
through $b$-tagging. From the above interpretation of these cuts, it
is likely that they are fairly complementary to $b$-tagging.
Assuming a single $b$-jet tagging efficiency of $40\%$ for the
signal, the other $60\%$ of the events passed by cuts (a--c) would
represent sizeable signal acceptance otherwise rejected by $b$-tagging.
D\O, on the other hand, could use these cuts alone to
match the previous background rejection of CDF, despite their
lack of a silicon vertex detector. Finally, we remark that although we
have discussed above only $m_t=175$ GeV, Table~1 shows that all of the
cuts have almost identical effect on a signal with $m_t=190$ GeV,
which reflects the relatively small dependence on the
exact value of $m_t$ for which the cuts were optimized.

We next present a new top quark reconstruction algorithm that,
applied to events passing the high acceptance cuts (a--b), can
measure $m_t$ directly.  Our first key observation is that the
measurement should be based on the hadronic decay $t \to b q \bar q$,
since the rather poor measurement of the neutrino momentum
significantly degrades the mass resolution for $t \to b \ell \nu$.
Our goal is to form a histogram of $m_{jjj}$ for  3-jet systems that
are tagged as coming from $t \to b q \bar q$,  using the $t \to b \ell
\nu$ mass only for the tagging,  {\it i.e.}, to recognize which of the
four jets came from the  leptonically decaying top, leaving the
leftover trio as the  hadronic decay.  The location of  the peak in
$m_{jjj}$ will measure $m_t$ (with Monte Carlo needed only  to assess
instrumental effects).  The backgrounds from QCD and from
incorrect jet assignment will be directly measured in a
model-independent way by fitting the histogram.  This is important
because it allows for the possibility that leading-order models of the
background such as {\sc vecbos} may be quite unreliable.

Unlike other
analyses, we do not attempt to fully reconstruct the event by trying
to identify which pair of the three jets in the hadronic decay came
{}from the $W$.  This keeps the ``combinatoric problem'' under control,
since it cuts down the possible jet assignments from 12 per event to
just 4. Also, since we treat the 3 jets in $t \to jjj$ symmetrically,
at the end of the analysis we can plot a histogram of dijet pair
masses from $t \to jjj$ candidates (3 combinations per event) and,
without reconstruction-induced bias, observe the $W
\to jj$ peak in it.  This will give an important independent
calibration of jet energy measurement and jet-finding algorithms.

Our partial reconstruction is carried out as follows.
For each event that passes the $S/B$ enhancement cuts (a--b), we
assign each of the four jets in turn to go with the lepton.
Let $m_{jjj}$ be the invariant mass of the remaining three jets.  We
select the assignment if
(1) $120 \, {\rm GeV} < m_{jjj} < 240  \, {\rm GeV}$;
(2) $|m_{j \ell \nu} - m_{\rm trial}| < 20 \, {\rm GeV}$;
(3) $|m_{j\ell\nu} - m_{\rm trial}|$ is the smallest of the four
possibilities that pass (1) and (2). We took the trial top quark mass
$m_{\rm trial}=175$ GeV, but show below that this choice affects only
the height, and not the location, of the mass peak in $m_{jjj}$. In
practice, of course, a range of $m_{\rm trial}$ may be swept to optimize
the signal peak. The mass range for $m_{jjj}$ is kept very broad, so
there is ample room to separate peak from background.  The mass range
for $m_{j \ell \nu}$ was chosen to keep $\sim 70 \%$ of the true signal.
We have checked that this algorithm does not produce fake peaks due to
either the QCD or combinatoric backgrounds.

Measurement of the neutrino momentum is crucial for measurement of
$m_{j \ell \nu}$.  The transverse momentum of $\nu$ is taken to be the
negative of the total $\vec{p}_T$ observed in the calorimeter, giving
it an uncertainty due to the uncertainties of all four jet $\vec{p}_T$'s
added in quadrature; plus contributions from
inaccurate measurement of the many low $p_T$ particles in the event,
the possibility of other neutrinos (e.g., from semi-leptonic decays in
one or both $b$-jets), and instrumental effects
due to gaps in the detector coverage.  The longitudinal momentum
of the neutrino can be computed from $m_{\ell \nu} = m_W$, with a
two-fold ambiguity in addition to uncertainties due to the width of
the $W$	 and the error in $\vec{p}_{T}^{\,\nu}$.  That computation is
usually expressed by a quadratic equation for $p_{L}^\nu$, but
it is much clearer to think of it as follows.  The invariant mass
$m_{\ell \nu}$ is given by
\begin{eqnarray}
m_{\ell \nu}^2 = 2 \, p_T^{\nu} \, p_T^{\ell} \,
[\cosh(\eta_\nu - \eta_\ell) - \cos(\phi_\nu - \phi_\ell)] \; .
\label{eq:planeq5}
\end{eqnarray}
By assuming $m_{\ell \nu} = m_W$ one determines
$\cosh(\eta_\nu - \eta_\ell)$ and hence $|\eta_\nu - \eta_\ell|$.
The two-fold solution ambiguity is due to the undetermined sign
of $\eta_\nu - \eta_\ell$:  {\it the two
solutions for $\eta_\nu$ lie on either side of $\eta_\ell$ and
equidistant from it.}  There will be considerable
uncertainty in $|\eta_\ell - \eta_\nu|$ due to errors in
$ p_T^{\nu}$ and $\phi_\nu$, the finite $W$ width, and because
$\cosh(\eta_\ell - \eta_\nu)$ is usually close to $1$,
where $m_{\ell \nu}$ is rather insensitive to $\eta_\ell - \eta_\nu$.
It can even happen ($\sim 20 \%$ of the time) that there is no
solution, in which case $\eta_\nu = \eta_\ell$ is the best guess.
When there are two solutions, we choose the sign of
$\eta_\ell - \eta_\nu$ to be that of $\eta_\ell$ (the solution with
the smaller $W$ energy), which most of the time is correct at the
Tevatron, since the $W$'s are produced rather centrally in rapidity
due to the limited total energy.  Even for the $\sim 22 \%$ of events
where the wrong solution is chosen, this rule is often adequate since
the two solutions are often close to each other,
since we only need the neutrino momentum to compute
$m_{b \ell \nu}$ which is not always very sensitive to $\eta_\nu$,
and since we only need $m_{b \ell \nu}$ measured accurately enough to
tag the correct one of the four jets.\footnote{
At much higher energies, such as at the LHC, there is no
clear way to choose the correct neutrino solution in order to evaluate
$m_{j \ell \nu}$.  For such a case, one can avoid choosing by
instead using $m_{j \ell \nu}^*$ which is defined by minimizing
$m_{j \ell \nu}$ with respect to $\eta_\nu$.  To an excellent degree
of approximation, that is equivalent to assigning $\eta_\nu$ to the
$p_T$ weighted average:
$\eta_\nu^* =
(p_T^\ell \eta_\ell + p_T^{j} \eta_{j})/(p_T^\ell + p_T^{j})$.
The ``Jacobian Peak'' in the amount of phase space
near the minimum causes a sharp peak in the probability distribution
for $m_{j \ell \nu}^*$ at a value only a slightly lower than the true
peak in $m_{j \ell \nu}$.  The quantity $m_{j \ell \nu}^*$ is analogous
to the ``transverse mass'' variable used in measuring $m_W$.}

Fig.~2 shows the resultant plot for $m_{jjj}$, with $m_t=175$,
$190$ GeV.
We have plotted only the events passing cuts (a--b) with
$|m_{j\ell\nu}-m_{\rm trial}| < 20$ GeV, $m_{\rm trial}=175$ GeV. This
includes 32.6/38.6 signal events for $m_t=175$ GeV, 18.2/25.5 signal
events for $m_t=190$ GeV, and only 13.7/19.8 background events.
The resulting clear peak has suffered almost no shift away from $m_t$,
despite simulated detector smearing effects and, importantly,
non-optimal
choice of $m_{\rm trial}$ in the case of $m_t=190$ GeV. (The peak for
$m_t=190$ GeV increases by $10\%$ if $m_{\rm trial}=190$ GeV is used,
but its
location is unchanged). This result
provides verification that our method, by relying totally on
$m_{j\ell\nu}$ for the trijet selection, avoids introducing bias
into the trijet mass.

A nice cross-check of a top quark peak found using this method is
shown in Fig.~3, where for each trio of jets in the peak,
each of the three dijet mass combinations is plotted (with weight
1/3 each).  A clear peak at $m_W$ appears, which will provide
a unique calibration for the hadronic calorimetry and the jet-finding
algorithm.   We also note that the combinatoric background under
the $W$ peak is substantial, which shows the wisdom of {\it not}
trying
to recognize $W \to jj$ as part of the $t \bar t$ event selection.

In conclusion, we have demonstrated the usefulness of the Binary
Decision Tree technique in separating signal and background events
for top quark production at the Tevatron. We showed why the new
observables $m_{jj}^6$, $m_{jjj}^4$, and $p_T^4(j)$ strongly enhance
the signal. We
introduced an algorithm to determine the top quark mass, which yields
a directly observable $t\to b q\bar{q}$ peak in a certain
$m_{jjj}$ distribution. We further showed $S/B \approx 4$ is
achievable (for $m_t=175$ GeV), with only about $50 \%$ loss of the
signal beyond typical minimal experimental acceptance cuts.

Finally, we point out that the methods derived here for
$t\bar t \to\ell + 4$
jets could be used in an analogous fashion to observe the total
hadronic signature $t\bar t \to 6$ jets\cite{giele}. We expect a
similar substantial increase of $S/B$ through {\sc hastac}-derived
cuts.  Given the higher event rate, further background suppression by
requiring one $b$-tagged jet would greatly reduce $B$ but leave
sufficient signal events. In analogy to the tag on $t\to b\ell\nu$,
we would pick from the 5 other jets the pair that 1) reconstructs a
$W$ boson and 2) best reconstructs $t\to jjj$ with the tagged $b$.
Then the invariant
mass of the {\it other} 3 jets should have an unbiased peak at $m_t$.
Work in this direction, as well as refinements of our method such as
inclusion of $b$-tagging information for $t\bar t\to\ell+4$ jets,
and consideration of events with less than $4$ observed jets is
in progress.

We thank the other {\sc hastac} collaborators, including R. Hatcher,
J. Linnemann, and in particular J. Hughes for much assistance in
implementing the current algorithm. S. Chao  provided crucial advice
on the optimization algorithm. We also had useful discussions with H.
Miettinen, H. Weerts and C.-P. Yuan.  P.A. was supported
in part by NSF grant PHY-9396022. D.B.-C. was supported in
part by NSF grant number PHY-9307980.

\newpage
\begin{figure}
\caption{ $m_{jj}^6$ in 2 GeV bins; the dashed curve is the
background, the solid curve the sum of signal ($m_t=175$ GeV) and
background, with acceptance cuts only at an integrated luminosity
of $\int {\cal L} dt =100$ pb$^{-1}$.
\label{dijet6}}
\end{figure}

\begin{figure}
\caption{ The reconstructed top quark mass, $m_{jjj}$, in 8
GeV bins, with cuts (a--b) and the requirement that
$|m_{j\ell\nu}-m_{\rm trial}| < 20$ GeV, for $m_{\rm trial}=175$ GeV.
The solid curve indicates the sum of signal, with $m_t=175$ GeV, and
background; the dashed curve gives the sum of signal, with $m_t=190$
GeV, and background; the dotted curve gives the background alone.  All
are presented for $\int {\cal L} dt =100$ pb$^{-1}$.
\label{trijet}}
\end{figure}

\begin{figure}
\caption{ Mass distribution (3 combinations per event) for dijets
formed from hadronically
decaying tops ($m_t=175$ GeV) identified using the top mass
reconstruction algorithm (cuts (a--b),
$|m_{j\ell\nu}-m_{\rm trial}| < 20$ with $m_{\rm trial}=175$ GeV, and
$|m_{jjj}-m_t| < 15$ GeV) at $\int {\cal L} dt =100$ pb$^{-1}$.
The $W$ boson mass peak, which is {\it not} used in the analysis,
shows up clearly.
\label{wmass}}
\end{figure}
%
\newpage
\widetext
\begin{table}
\renewcommand{\arraystretch}{1.0}
\caption{Effect of {\sc hastac}-derived generalized cuts on signal and
background, in events at $\int {\cal L} dt =100$ pb$^{-1}$.
The percentage of events relative to acceptance cuts only
is given in parentheses.}
\begin{tabular}{ccccc}
& Cuts & Signal ($m_t=175$ GeV) & Signal ($m_t=190$ GeV) & Background\\
\tableline
& Acceptance Cuts Only & 49.0 (100\%) & 32.3 (100\%) & 116.0 (100\%) \\
& (a)                  & 40.8 (83\%)  & 28.5 (88\%)  &  29.2  (25\%) \\
&(a--b)                & 38.6 (79\%)  & 25.5 (79\%)  &  19.8  (17\%) \\
&(a--b) and (c)        & 30.0 (61\%)  & 20.4 (63\%)  &  10.0   (9\%) \\
&(a--b) and (d)        & 22.4 (46\%)  & 16.5 (51\%)  &   6.0   (5\%) \\
\tableline
(a) &\multicolumn{4}{c}{$(0.76 \, p_T^4(j) + 0.14 \, m_{jjj}^4 + 0.65
\, m_{jj}^6 + 0.13  \, p_T(W) )/m_W > 1.0 $ } \\
(b) & \multicolumn{4}{c}{$0.43 \, m_{jjj}^4/m_{W} - 0.32 \,
m_{jj}^6/m_W +  0.089 \, |\eta^4(j)| + 0.097 \,|\eta^3(j)| < 1.0$ } \\
(c) & \multicolumn{4}{c}{$( - 0.13 \, m_{jjj}^1 + 0.79
\,m_{jjj}^4)/m_W > 1.0 $ } \\
(d) & \multicolumn{4}{c}
{$- 0.31 \, m_{jjj}^1/m_W + 1.4 \,m_{jjj}^4/m_W -
0.63 \,|\eta^4(j)| > 1.0$} \\
\end{tabular}
\label{table1}
\end{table}
\narrowtext

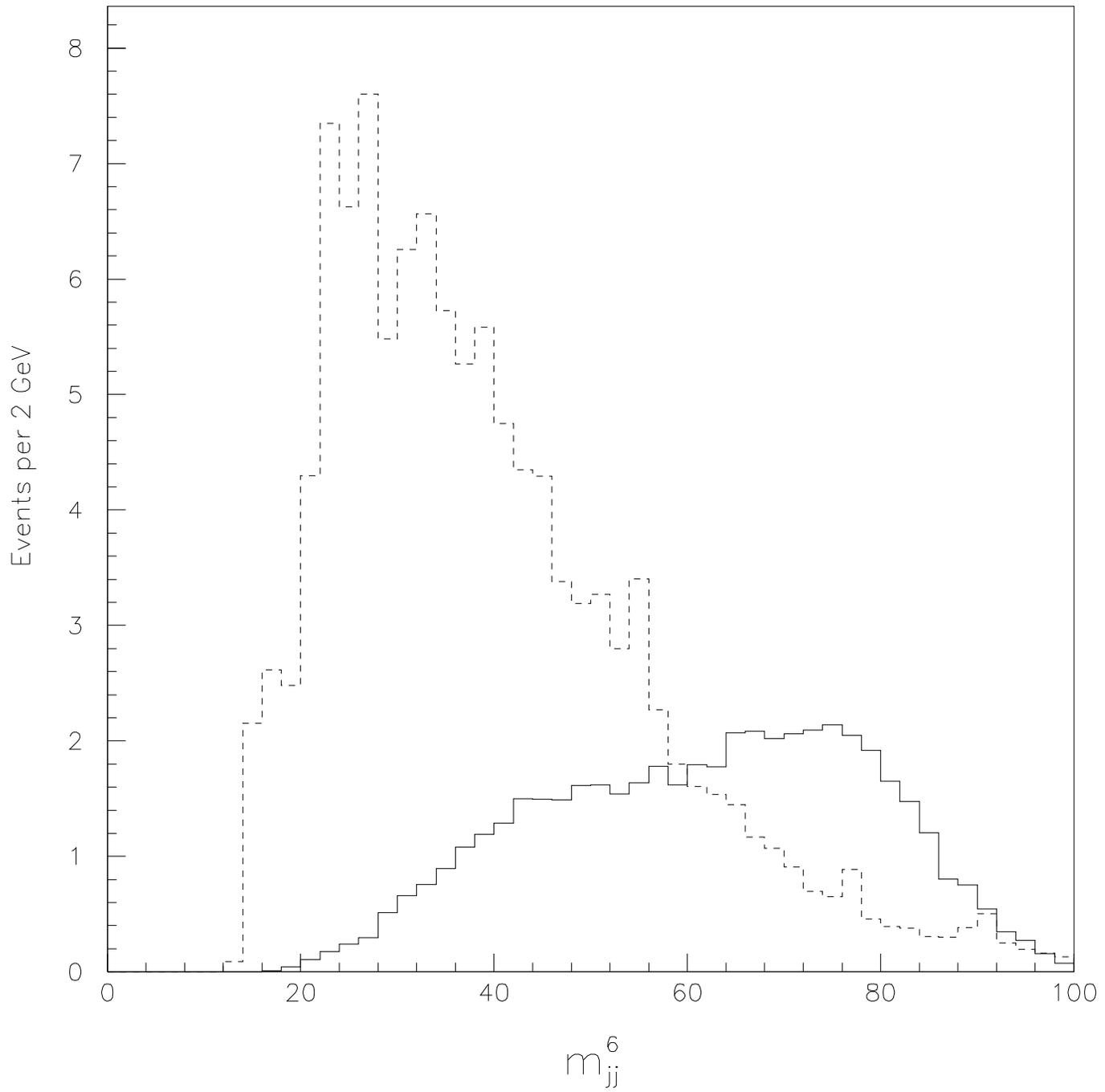
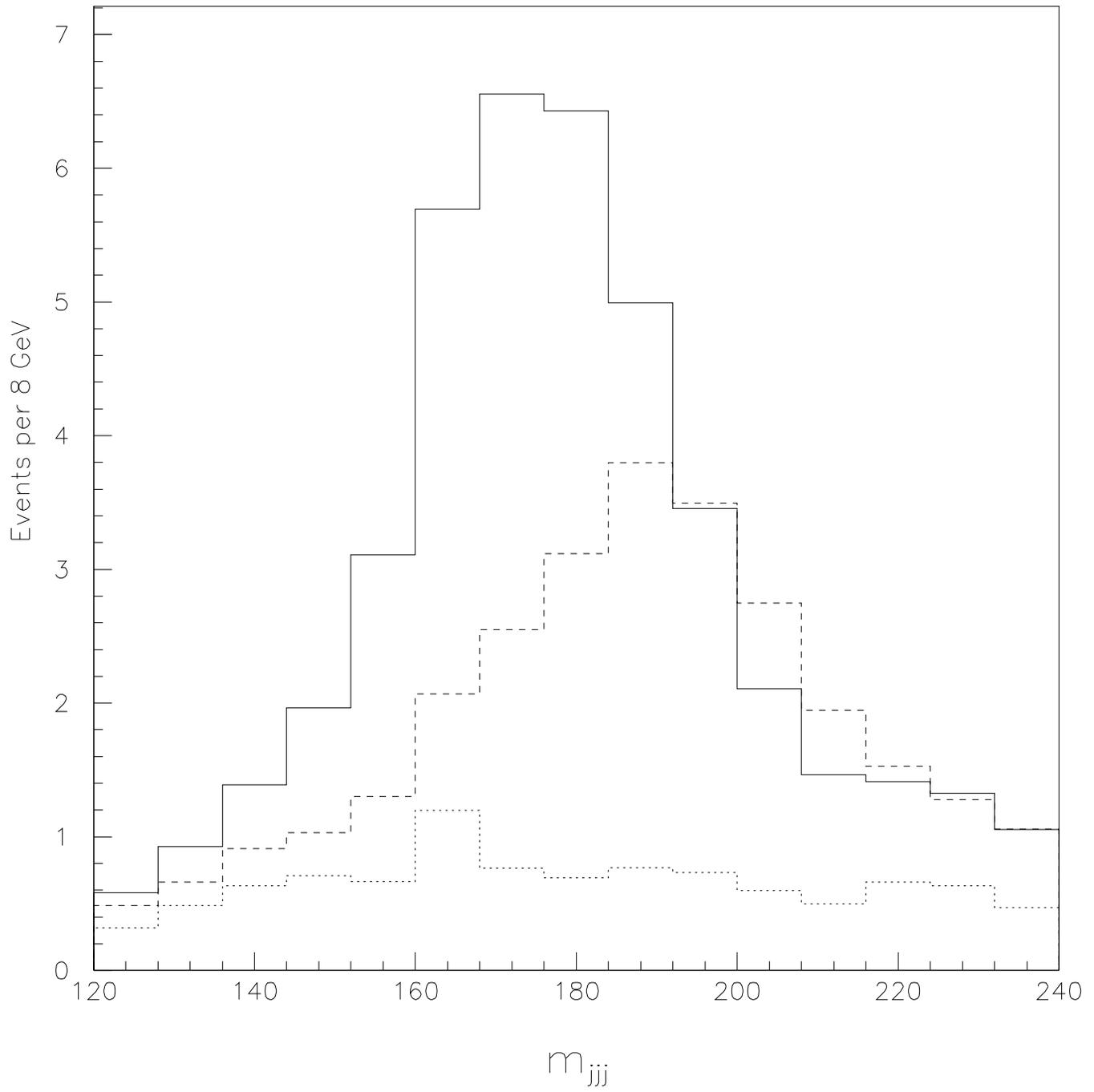
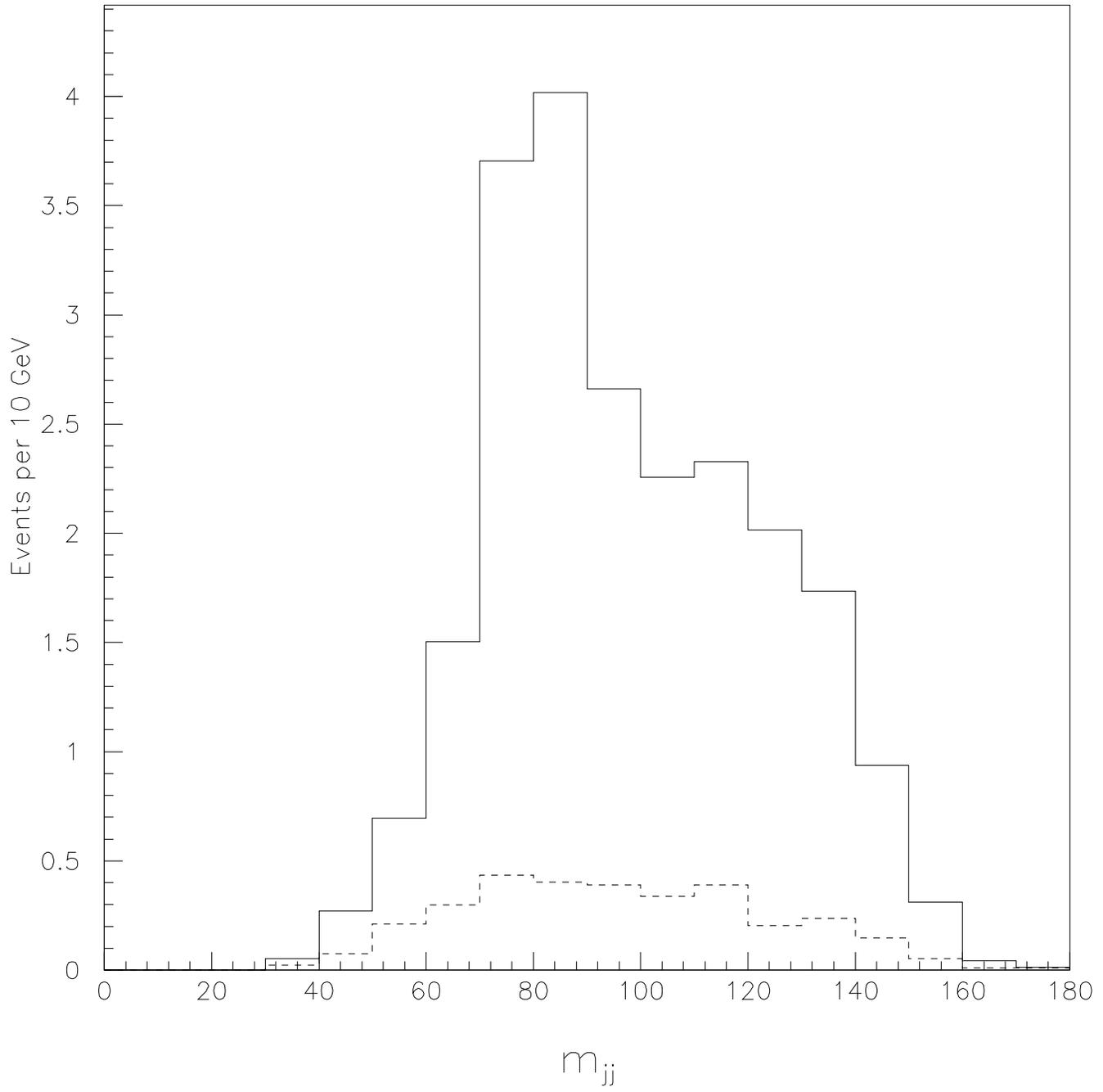
\end{document}